\documentclass[11pt,a4paper,english,superscriptaddress]{revtex4}
\usepackage{amssymb}
\usepackage{amsmath} 
\usepackage{slashed}
\usepackage{graphicx}
\usepackage{extpfeil}
\usepackage{babel}
\usepackage{hyperref}
\usepackage{color}
\usepackage{comment}
\usepackage[normalem]{ulem}
\usepackage{cancel}

\makeatletter\AtBeginDocument{\let\@elt\relax}\makeatother

\begin{document}

\title{On the physical running of the electric charge in a dimensionless theory of gravity}

\author{M. Gomes}
\email{mgomes@fma.if.usp.br}
\affiliation{Instituto de F\'\i sica, Universidade de S\~ao Paulo \\
Caixa Postal 66318, 05315-970, S\~ao Paulo, SP, Brazil}
	
\author{A.~C.~Lehum}
\email{lehum@ufpa.br}
\affiliation{Faculdade de F\'isica, Universidade Federal do Par\'a, 66075-110, Bel\'em, Par\'a, Brazil.}

\author{A. J. da Silva}
\email{ajsilva@fma.if.usp.br}
\affiliation{Instituto de F\'\i sica, Universidade de S\~ao Paulo \\
Caixa Postal 66318, 05315-970, S\~ao Paulo, SP, Brazil}

\begin{abstract}
We revisit the renormalization of the gauge coupling in massless QED coupled to a scaleless quadratic theory of gravity. We compare two alternative prescriptions for the running of the electric charge: (i) the conventional $\mu$-running in minimal subtraction, and (ii) a ``physical'' running extracted from the logarithmic dependence of amplitudes on a hard scale $Q^{2}$ (e.g., $p^{2}$ or a Mandelstam invariant) after removing IR effects. At one loop, using dimensional regularization with an IR mass regulator $m$, we compute the photon vacuum polarization. We find a clean separation between UV and soft logarithms: the former is gauge and process independent and fixes the beta function, whereas the latter encodes nonlocal, IR-dominated contributions that may depend on gauge parameters and must not be interpreted as UV running. In the quadratic-gravity sector, the photon self-energy is UV finite—the $\ln\mu^{2}$ pieces cancel—leaving only $\ln(Q^{2}/m^{2})$ soft logs. Consequently, quadratic gravity does not modify the one-loop UV coefficient and thus does not alter $\beta(e)$. Therefore, the ``physical'' running coincides with the $\mu$-running in QED at one loop. Our analysis clarifies how to extract a gauge and process independent running in the presence of gravitational interactions and why soft logs from quadratic gravity should not contribute to $\beta(e)$.
\end{abstract}


\maketitle

\section{Introduction}

Even though Einstein's theory of gravity, quantized for small fluctuations around a flat metric, is famously nonrenormalizable~\cite{'tHooft:1974bx,PhysRevLett.32.245,Deser:1974cy}, it can be treated as an effective field theory~\cite{Donoghue:1994dn,Burgess:2003,Donoghue:2012zc,Donoghue:2017pgk}. This approach allows for predictions of processes below the Planck scale, such as quantum corrections to the Newtonian potential~\cite{Donoghue:1994dn,Faller:2007sy}, Schwarzschild solution~\cite{Bjerrum-Bohr:2014zsa}, and FLRW metrics~\cite{Donoghue:2014yha}. Nevertheless, a quantum description of gravity at and above the Planck scale remains an unresolved problem. Various approaches have been proposed to formulate a quantum theory of gravity applicable at energy scales beyond the Planck scale, including Loop Quantum Gravity~\cite{Rovelli:1997yv}, the Functional Renormalization Group~\cite{Percacci:2007sz}, and Lattice formulations~\cite{Hamber:2009mt}, superstrings \cite{polchinski} and causal sets \cite{Surya:2019ndm}.

An additional and relatively conservative approach involves the inclusion of quadratic terms in the curvature tensors~\cite{Stelle:1976gc,Avramidi:1985ki,deBerredo-Peixoto:2003jda,Narain:2013eea,Salvio:2014soa,Alvarez-Gaume:2015rwa,Salvio:2017qkx}, which improves the UV behavior of the theory, rendering it perturbatively renormalizable~\cite{Stelle:1976gc,Narain:2013eea}. This improvement arises because the graviton propagator in such theories behaves as $1/p^4$, in contrast to the usual $1/p^2$ behavior in Einstein's gravity. At the same time, quadratic gravity may face challenges related to unitarity due to the appearance of ghostlike degrees of freedom~\cite{Salvio:2014soa,Alvarez-Gaume:2015rwa,Salvio:2017qkx,Holdom:2015kbf,Anselmi:2018ibi,Salvio:2018crh,Salvio:2018kwh,Donoghue:2021cza,Buccio:2024hys}. Nevertheless, it has been argued that scattering amplitudes can remain well behaved in the ultra-Planckian regime~\cite{Holdom:2021oii,Holdom:2021hlo,Cunha:2025djg}. As emphasized in Ref.~\cite{Donoghue:2021cza}, these issues merit further scrutiny and are not conclusively settled in the literature. The viability of quadratic gravity as a realistic framework for quantum gravity therefore remains an open question requiring additional investigation.

One of the key areas of interest in quadratic theories of gravity is the study of renormalization group functions, particularly the beta functions of the coupling constants. Recently, some authors have raised questions about earlier results for the beta functions of the gravitational couplings~\cite{Buccio:2024hys,Buccio:2024omv}. Their criticism pertains to the procedure used to compute the beta functions, which is a fundamental step in determining the running coupling constants. In their analysis, the authors of Refs.~\cite{Buccio:2024hys,Buccio:2024omv} compared the beta functions of the gravitational constants obtained using the minimal subtraction (MS) renormalization scheme, referred to as \textit{$\mu$-running}, with those derived from the dependence on the external momentum $p$, through $\ln(-p^2)$, referred to as \textit{physical running} (p-running). They concluded that the beta functions derived from these two distinct prescriptions differ from each other, arguing that the physically meaningful running coupling constants should be determined from the $\ln(-p^2)$ dependence, i.e., the physical running.

Previous analyses have provided valuable insights, but were carried out in pure gravity and reported results in a specific gravitational gauge, under the expectation that physical observables are ultimately gauge independent. However, subsequent work has shown that the “physical running” as proposed in Refs.~\cite{Buccio:2024hys,Buccio:2024omv} is gauge dependent~\cite{Buccio:2025tci,Salvio:2025cmi}, and that scattering amplitudes can be used to extract a gauge–independent notion of running~\cite{Salvio:2025cmi}.

Here we extend the discussion to massless QED coupled to a quadratic theory of gravity and compare two prescriptions for the running of the electric charge—the conventional $\mu$-running and the “physical” running—while allowing for a more general gravitational gauge. Our main result is that the two prescriptions are equivalent once UV and IR effects are cleanly separated.

The paper is organized as follows: In Sec.~II, we present the action of the model, derive the Lagrangian for small fluctuations around a flat metric, and obtain the propagators of the model. In Sec.~III, we evaluate the one-loop self-energy of the photon, we discuss the determination of the beta function of the electric charge using two different prescriptions, namely the {$\mu$-running} and the {physical running}, we compute the gravitational corrections to the photon polarization tensor, demonstrating that the two different prescriptions lead to the same result for the beta function of the electric charge after removing IR effects. Finally, in Sec.~IV, we present our concluding remarks. Throughout this paper, we adopt natural units, $c = \hbar = 1$, and use the metric signature $(+---)$.

\section{The QED agravity Lagrangian}\label{sec11}

Consider the model described by the action
\begin{eqnarray}\label{Agravity-QED}
	\mathcal{S} = \int d^4x \sqrt{-g} \left[ \frac{R^2}{6f_0^2} + \frac{1}{f_2^2} \left(\frac{1}{3} R^2 - R^{\mu\nu} R_{\mu\nu}\right) 
	-\frac{1}{4} g^{\mu\nu} g^{\alpha\beta} F_{\mu\alpha} F_{\nu\beta} + i\, \bar{\psi} \left(\frac12\overleftrightarrow\nabla_{\mu} - ig A_\mu \right) \gamma^\mu \psi \right],
\end{eqnarray}
\noindent
where $R$ and $R_{\mu\nu}$ denote the Ricci scalar and Ricci tensor, respectively. We adopt the notation $\bar{\psi}\overleftrightarrow\nabla_{\mu} \psi = \bar{\psi}\overrightarrow\nabla_{\mu}\psi-\bar{\psi}\overleftarrow\nabla_\mu\psi$ where $\nabla_\mu$ is the spinor covariant derivative. The Maxwell's field strength is given by $F_{\mu\nu} = \partial_\mu A_\nu - \partial_\nu A_\mu$.

The Dirac matrices are defined in terms of the vierbein by $\gamma^\mu = \gamma^\alpha e^\mu_\alpha$, where the spacetime metric satisfies $g_{\mu\nu} = e_\mu^\alpha e_\nu^\beta \eta_{\alpha\beta}$. The spinor covariant derivatives are defined as $\overrightarrow{\nabla}_\mu \psi = (\partial_\mu + i\omega_\mu)\psi$ and $\bar{\psi} \overleftarrow{\nabla}_\mu = \partial_\mu \bar{\psi} - i \bar{\psi} \omega_\mu$, where the spin connection is
\begin{eqnarray}
\omega_\mu = \frac{1}{4} \sigma^{\alpha\beta} \left[ e^\nu_\alpha (\partial_\mu e_{\beta\nu} - \partial_\nu e_{\beta\mu}) 
+ \frac{1}{2} e^\rho_\alpha e^\sigma_\beta (\partial_\sigma e_{\gamma\rho} - \partial_\rho e_{\gamma\sigma}) e^\gamma_\mu 
- (\alpha \leftrightarrow \beta) \right],
\end{eqnarray}
with $\sigma^{\alpha\beta} = \frac{i}{2}[\gamma^\alpha, \gamma^\beta]$. Throughout, Greek indices from the middle of the alphabet ($\mu, \nu, \dots$) refer to general spacetime coordinates, while those from the beginning of the alphabet ($\alpha, \beta, \dots$) denote locally inertial (tangent space) coordinates.

In order to study the model, we expand the spacetime metric around the flat Minkowski background as
\begin{equation}\label{metric} 	
g_{\mu\nu} = \eta_{\mu\nu} + h_{\mu\nu} \quad \text{(exact)}, \qquad g^{\mu\nu} = \eta^{\mu\nu} - h^{\mu\nu} + \cdots,
\end{equation}
where spacetime indices (Greek letters) are raised and lowered using the flat Minkowski metric $\eta_{\mu\nu} = \mathrm{diag}(+,-,-,-)$. Within the one-graviton exchange approximation, the effective Lagrangian ${\cal L}$ can be organized into three basic contributions,  
\begin{eqnarray}
{\cal L}= {\cal L}_{h}^{0}+ {\cal L}_{f}+ {\cal L}_{A}, 
\end{eqnarray}
where ${\cal L}_{h}^{0}$ denotes the gravitational sector without self-interaction terms, ${\cal L}_{f}$ corresponds to the fermion sector, and ${\cal L}_{A}$ describes the gauge sector. Explicitly, with $h=h^{\mu}_{\mu}$, we have
\begin{eqnarray}\label{Lg_quadratic}
	\mathcal{L}_h^0 &=& 
	-\frac{1}{8 f_2^2}	\Big[ 
	(\partial^\sigma\partial^\mu-\Box \eta^{\mu\sigma})h_{\mu\nu} (\partial^\rho\partial^\nu-\Box \eta^{\nu\rho})h_{\rho\sigma}
	+(\partial^\rho\partial^\mu-\Box \eta^{\mu\rho})h_{\mu\nu} (\partial^\sigma\partial^\nu-\Box \eta^{\nu\sigma})h_{\rho\sigma}\nonumber\\
	&& 
	-\frac{2}{3}(\partial^\nu\partial^\mu-\Box \eta^{\mu\nu})h_{\mu\nu} (\partial^\rho\partial^\sigma-\Box \eta^{\sigma\rho})h_{\rho\sigma}\Big]
	+\frac{1}{6f_0^2} \left( \partial^{\mu}\partial^\nu h_{\mu\nu}-\Box h \right)^2, 
\end{eqnarray}
for the gravitational sector,
\begin{subequations}\label{Lfermions}
\begin{eqnarray}
  \mathcal{L}_f &=& \mathcal{L}_f^0 + e\,\bar{\psi}\gamma^\mu A_\mu \psi  
  -\frac{1}{2} e \left(h\,\eta_{\mu\nu} - h_{\mu\nu}\right)\bar\psi \gamma^\mu A^\nu \psi
  + \mathcal{L}_f^1 + \cdots, \\
  \mathcal{L}_f^0 &=& \frac{i}{2}\left(\bar{\psi}\gamma^\mu\partial_\mu\psi - \partial_\mu\bar{\psi}\gamma^\mu\psi\right),\\
  \mathcal{L}_f^1 &=& \frac{1}{2}h\,\mathcal{L}_f^0 - \frac{i}{4}h_{\mu\nu}\left(\bar{\psi} \gamma^\mu\partial^\nu\psi - \partial^\nu\bar{\psi} \gamma^\mu\psi\right),
\end{eqnarray}
\end{subequations}
for the fermionic fields, and
\begin{subequations}\label{LA}
 \begin{eqnarray}
  \mathcal{L}_A &=& \mathcal{L}_A^0 + \mathcal{L}_A^1 + \mathcal{L}_A^2 +\cdots,\\
  \mathcal{L}_A^0 &=& -\frac{1}{4}F_{\mu\nu}F^{\mu\nu}, \\
  \mathcal{L}_A^1 &=& \frac{1}{2}h^\tau_{~\nu}F^{\mu\nu} F_{\mu\tau} + \frac{1}{2}h\,\mathcal{L}_A^0,\\
  \mathcal{L}_A^2 &=& \frac{1}{8}\left(h^2-2 h^{\mu\nu}h_{\mu\nu} \right)\mathcal{L}_A^0 + \frac{1}{4}F_{\alpha\beta} F_{\rho\sigma} \left(h\,h^{\alpha\rho}\eta^{\beta\sigma}-2h^{\alpha}_{\mu}h^{\mu\rho}\eta^{\beta\sigma} -h^{\alpha\rho}h^{\beta\sigma}\right),
 \end{eqnarray}
\end{subequations}
for the gauge fields. A detailed expansion of the interaction terms contained in Eqs.~\eqref{Lfermions} and \eqref{LA} can be found in Ref.~\cite{Choi:1994ax}. 


In addition to Eq.\eqref{Agravity-QED} expanded around flat metric, we must include the gauge-fixing Lagrangian, given by
\begin{eqnarray}
	\mathcal{L}_{GF} = -\frac{1}{2\xi_g}f^\mu\Box f_\mu - \frac{1}{2\xi_a}(\partial^\mu A_\mu)^2 + \mathcal{L}_{ghosts},
\end{eqnarray}
\noindent where $f_\mu = \partial^\nu(h_{\mu\nu} - \frac{1}{2}\eta_{\mu\nu}h)$ and $\mathcal{L}_{ghosts}$ represents the ghosts Lagrangian, which we will omit here as it does not contribute to the calculations at the order we are working, i.e., the one-graviton exchange approximation.

The quadratic part of the action yields the subsequent propagators:
\begin{eqnarray}
S_F(p) &=& \frac{i\slashed{p}}{p^2},\nonumber\\
\Delta^{\mu\nu}(p) &=& -\frac{i}{p^2}\left[\eta^{\mu\nu}-\left(1-\xi_a \right)\frac{p^\mu p^\nu}{p^2} \right],\nonumber\\
\Delta_{\mu\nu\rho\sigma}(p) &=& \frac{i}{p^4}\left[-2f_2^2 P^{(2)}_{\mu\nu\rho\sigma} +f_0^2 P^{(0)}_{\mu\nu\rho\sigma}+2\xi_g\left(P^{(1)}_{\mu\nu\rho\sigma} +\frac{1}{2} P^{(0w)}_{\mu\nu\rho\sigma}\right) \right],
\end{eqnarray}
\noindent where $S_F(p)$, $\Delta^{\mu\nu}(p)$ and $\Delta_{\mu\nu\rho\sigma}(p)$ are the fermion, photon and graviton propagators, respectively. The projectors are defined as follows
\begin{eqnarray}
P^{(2)}_{\mu\nu\rho\sigma} &=& \frac{1}{2}T_{\mu\rho}T_{\nu\sigma} +\frac{1}{2}T_{\mu\sigma}T_{\nu\rho}-\frac{1}{D-1}T_{\mu\nu}T_{\sigma\rho},\nonumber\\
P^{(1)}_{\mu\nu\rho\sigma} &=& \frac{1}{2}\left( T_{\mu\rho}L_{\nu\sigma} +T_{\mu\sigma}L_{\nu\rho} + L_{\mu\rho}T_{\nu\sigma} + L_{\mu\sigma}T_{\nu\rho} \right), \nonumber\\
P^{(0)}_{\mu\nu\rho\sigma} &=& \frac{1}{D-1}T_{\mu\nu}T_{\sigma\rho},\nonumber\\
P^{(0w)}_{\mu\nu\rho\sigma} &=& L_{\mu\nu}L_{\sigma\rho},
\end{eqnarray}
\noindent with
\begin{eqnarray}
T_{\mu\nu} &=& \eta_{\mu\nu}-\frac{p_\mu p_\nu}{p^2},\nonumber\\
L_{\mu\nu} &=& \frac{p_\mu p_\nu}{p^2}.
\end{eqnarray}

With the propagators of the model established, we now proceed in Sec.~III to evaluate the one-loop corrections to the photon self-energy, which serve as the basis for exploring the renormalization of the electric charge.

\section{The electric charge beta function}

\subsection{The one loop photon self-energy}

\begin{figure}[ht]
	\includegraphics[angle=0 ,width=15cm]{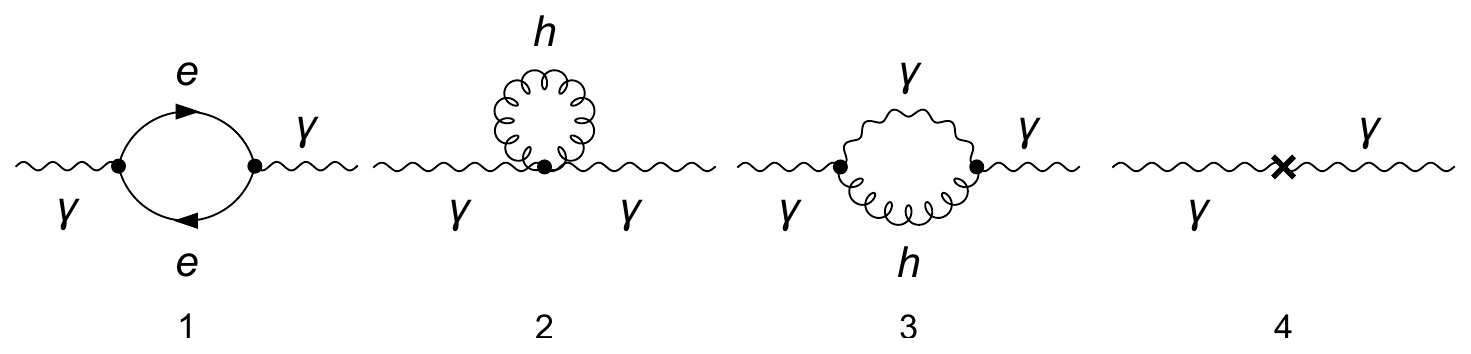}
	\caption{Photon field self-energy. Wavy,  continuos and wiggly lines represent the photon, fermionic and graviton propagators, respectively}
	\label{fig01}
\end{figure}

To begin, we will evaluate the one-loop corrections to the photon self-energy. The relevant Feynman diagrams are illustrated in Figure \ref{fig01}. To manage IR divergences, we adopt a straightforward prescription for all the propagators, modifying the expression $\dfrac{1}{k^2}$ to $\dfrac{1}{k^2 - m^2}$, where $m$ acts as an IR regulator. The calculation of the one-loop amplitudes has been implemented using a set of Mathematica\textsuperscript{TM} packages \cite{feyncalc,feynarts,feynrules,feynhelpers}.

The first diagram corresponds to the standard process in QED without the influence of gravitational interactions. The expression for the diagram depicted in Figure \ref{fig01}.1 is given by
\begin{eqnarray}
\Pi^{\mu\nu}_1(p) = 
\frac{2(D-2) i \pi^{2}\,e^{2}}{(D-1)\,p^{2}}\;
\big(\eta^{\mu\nu}\,p^{2}-p^{\mu}p^{\nu}\big)\,
\Big[ 2~A_{0}(m^{2})
-\left(\frac{4 m^{2}}{(D-2)}+p^{2}\right)\,B_{0}\!\big(p^{2},m^{2},m^{2}\big)\Big],
\end{eqnarray}
where the integrals $A_0$ and $B_0$ are defined in the Appendix.

Adding this contribution to the counterterm diagram shown in Fig.~\ref{fig01}.4, and defining the vacuum–polarization tensor by
\begin{equation}
\Pi^{\mu\nu}(p) \equiv \big(p^{2}\eta^{\mu\nu}-p^{\mu}p^{\nu}\big)\,\Pi(p),
\end{equation}
the large–momentum expansion for \(p^{2}\gg m^{2}\) yields
\begin{eqnarray}\label{piQED}
\Pi_{\text{QED}}(p)
&=&
\frac{2 i \pi^{2} (D-2)\,e^{2}}{(2\pi)^{D/2}(D-1)\,p^{2}}\,
\Big[2\,A_{0}(m^{2})
-\left(\frac{4 m^{2}}{(D-2)} +p^{2}\right)\,B_{0}\!\big(p^{2},m^{2},m^{2}\big)\Big]
- i\,\delta_{3}
\nonumber\\[2mm]
&=&
-\frac{i\,e^{2}}{12\,\pi^{2}}
\left[\frac{1}{\epsilon}+ \ln\left(\frac{-p^{2}}{4\pi\,\mu^{2}\,\mathrm{e}^{5/3-\gamma_E}}\right)\right]
- i\,\delta_{3}
+\mathcal{O}\!\left(\frac{m^{2}}{p^{2}}\right),
\label{pisqed}
\end{eqnarray}
where $\delta_{3}\equiv (Z_{3}-1)$ is the photon wave–function counterterm, $\epsilon=(D-4)/2$ is the UV regulator, and $A_{0}$ and $B_{0}$ denote the standard scalar one and two–point Passarino–Veltman integrals~\cite{tHooft:1978jhc}, respectively. The finite terms contain dependence on the logarithmic of $p^{2}/m^2$, as detailed in the Appendix. 

It is worth emphasizing that $\ln m^{2}$ terms arising from $A_{0}$ and $B_{0}$ cancel against each other. Consequently, in QED the photon vacuum polarization carries no soft IR logarithm of the form $\ln(p^{2}/m^{2})$; the only large $p^{2}$ logarithm is the UV $\ln(p^{2}/\mu^{2})$.

We are now prepared to discuss the distinction between $\mu$-running and the physical running~\cite{Buccio:2024hys,Buccio:2024omv}. The $\mu$-running refers to the computation of the beta function of the electric charge from the UV divergent part of the counterterm, as obtained using the minimal subtraction (MS) scheme of renormalization. In contrast, the {physical} running is characterized by the logarithmic dependence on the external momentum $p^2$, which results from the counterterm's dependence on the renormalization scale $M$, with $p^2 = -M^2$ as the chosen renormalization point.

\subsection{$\mu$-running versus p-running in QED}


First, we will derive the beta function $\beta_\mu(e) \equiv \mu \dfrac{de}{d\mu}$. By renormalizing Eq.\eqref{piQED} through MS prescription, we determine the counterterm as $\delta_3=-\frac{i\,e^{2}}{12\,\pi^{2}\epsilon}$ and the renormalized vacuum-polarization function in the high-energy regime results in
\begin{eqnarray}
\Pi_{\text{QED}}(p)
&=&-
\frac{i\,e^{2}}{12\,\pi^{2}}
 \ln\left(\frac{-p^{2}}{4\pi\,\mu^{2}\, \mathrm{e}^{5/3-\gamma_E}}\right).
\label{pisqed}
\end{eqnarray}

In dimensional regularization, the relation between the renormalized and bare charges reads
\begin{eqnarray}
Z_1\, e\,\mu^{\epsilon}
=
Z_2\, Z_3^{1/2}\, e_0 .
\end{eqnarray}
Using the Ward identity $Z_1=Z_2$, we obtain
\begin{eqnarray}
e\,\mu^{\epsilon}=Z_3^{1/2}\,e_0=\left(1+\frac{1}{2}\delta_3\right)\,e_0 .
\end{eqnarray}
Keeping $e_0$ fixed and differentiating with respect to $\ln\mu$, the one-loop result becomes
\begin{eqnarray}\label{beta_e_qed1l}
\beta_\mu(e)
=
\lim_{\epsilon\to 0}\, \mu\frac{de}{d\mu}
=
\frac{e^3}{12\pi^2}.
\end{eqnarray}

On the other hand, we can also renormalize the photon self-energy by imposing a different renormalization procedure. Following the prescription of Peskin and Schroeder \cite{Peskin:1995ev}, we impose the condition
\begin{eqnarray}
\Pi(p)\Big{|}_{p^2=-M^2}=0,
\end{eqnarray}
\noindent where the notation $\Big{|}_{p^2=-M^2}$ indicates evaluation at
$p^2=-M^2$. This choice yields the counterterm
\begin{eqnarray}
\bar{\delta}_3 &=&
-\frac{e^{2}}{12\,\pi^{2}} \left[ \frac{1}{\epsilon}
+ \ln\left(\frac{M^{2}}{4\pi\,\mu^{2}\, \mathrm{e}^{5/3-\gamma_E}}\right)\right],
\end{eqnarray}
\noindent where the explicit dependence on $\ln(M^2)$ originates from the scalar integral $B_0$ evaluated at $p^2=-M^2$. The renormalization scale $M$ is taken to be much larger than the IR regulator $m$. The renormalized polarization function then
takes the form
\begin{eqnarray}
\Pi_{\text{QED}}(p)
&=&
-\frac{i\,e^{2}}{12\,\pi^{2}}
 \ln\left(\frac{-p^{2}}{M^{2}}\right).
\end{eqnarray}

It is noteworthy that, considering the relation $e = Z_3^{1/2} e_0 = \left(1 + \dfrac{\bar\delta_3}{2}\right)e_0$, we can compute the beta function of the electric charge as it depends on the renormalization scale $M$, which is closely related to the external momentum $p$ dependence of the one-loop amplitude:
\begin{eqnarray}\label{beta_e_qed1l_physical}
	\beta_M(e) = M \frac{de}{dM} = \frac{e^3}{12\pi^2}.
\end{eqnarray}
 
It is important to note that both procedures for extracting the beta function yield the same result for $\beta(e)$ in QED. In the next section we examine how quadratic–gravity corrections can spoil this equivalence, and we show how UV poles provide an unambiguous diagnostic for isolating the physical running.

\subsection{Gravitational corrections}

In this section we compute the one–loop gravitational contribution to the photon vacuum polarization. The relevant Feynman diagrams are shown in Figs.~\ref{fig01}.2 and \ref{fig01}.3. For the topology displayed in these figures, the contribution reads
\begin{eqnarray}
\Pi_{2+3}(p)
&=&
-\frac{i\,\pi^{2}}{6}\,
\frac{\bigl[B_{0}(0,\,m^{2},m^{2})-B_{0}(p^{2},\,m^{2},m^{2})\bigr]\,
      \bigl[m^{2}\,(f_{2}^{2}+\xi_{g})-(3f_{2}^{2}-\xi_{g})\,p^{2}\bigr]}
     {(2\pi)^{D/2}p^{2}}+\cdots
\nonumber\\
&=&
\frac{i}{576\,\pi^{2}}\left[3\left(f_{0}^{2} - 7 f_{2}^{2} + 18\,\xi_g\right)
  - 2\left(f_{0}^{2} - 10 f_{2}^{2} + 9\,\xi_g\right)
    \ln\!\left(-\frac{p^{2}}{m^{2}}\right)
\right]+\mathcal{O}\left(\frac{m^2}{p^2}\right),
\label{gravPi23}
\end{eqnarray}
where $m$ denotes the IR mass regulator and $\xi_{g}$ is the gravitational gauge parameter. The ellipsis $(\cdots)$ indicates the term proportional to UV–finite scalar integrals--$C_{0}$, $D_{0}$ and $E_{0}$--which are omitted for brevity; being UV–finite, it does not affect the running of the coupling constant~\cite{Donoghue:2024uay}. The second line follows from the large–momentum expansion $p^{2} \gg m^{2}$ (here the finite contributions not explicitly shown in the previous line have been taken into account). We also note the appearance of an IR logarithm, whose interpretation and cancellation are discussed below.

The above gravitational contributions yields a UV–finite result. Consequently, no gravitational correction to the one–loop QED beta function $\beta(e)=\mu\,de/d\mu$ is expected, in agreement with Refs.~\cite{Narain:2013eea,Salvio:2014soa}. Therefore, the $\mu$–running of the electric charge at one loop coincides with the case without gravitational interactions.

On the other hand, Eq.~\eqref{gravPi23} contains a $\ln(-p^{2})$ term originated from $B_{0}(p^{2},m^{2},m^{2})$ that is explicitly $\xi_{g}$–dependent. Although such a momentum logarithm might be mistaken for a running effect, the $\ln(\mu^{2})$ pieces cancel within the combination $B_{0}(0,m^{2},m^{2})-B_{0}(p^{2},m^{2},m^{2})$, leaving only soft IR logarithms governed by the regulator $m$, namely $\ln m^{2}$ and $\ln(-p^{2}/m^{2})$. These terms are nonlocal, IR–dominated (and gauge–parameter dependent), and therefore must not be interpreted as contributions to the UV, physical running of the electric charge or to $\beta(e)$. 

Moreover, logarithms of the same IR type, such as $\ln(-p^{2}/m^{2})$, also arise from UV–finite integrals (e.g.\ the scalar integrals $C_{0}$, $D_{0}$, $E_{0}$, $\ldots$). These contributions encode soft/collinear dynamics regulated by $m$ and, while relevant for the full IR structure and its cancellation against real emission in inclusive observables~\cite{Zwanziger:1974jz} (see also M.~D.~Schwartz~\cite{Schwartz:2014sze}, Chap.~20 and references therein), they must be discarded when isolating the UV coefficient of $\ln(p^{2}/\mu^{2})$ (or $\ln(p^{2}/M^{2})$) that determines the physical running.

\section{Final Remarks}\label{summary}

In this work we have computed the beta function of the electric charge using two complementary prescriptions:
(i) the conventional $\mu$-running, defined by the \(\mu\)-dependence of renormalized parameters, and
(ii) the ``physical'' running, inferred from the dependence of amplitudes on a external momentum scale $M$ (e.g. a Mandelstam invariant or $p^{2}$ in a two–point function), after IR effects been consistently removed. Concretely, we evaluated the photon vacuum polarization at one loop and extracted the wave–function counterterm in both setups.

In pure QED the two procedures are equivalent and yield the same result for $\beta(e)$.
When quadratic–gravity effects are included, however, the appearance of additional logarithms can obscure this equivalence unless one carefully distinguishes
\emph{UV} from \emph{soft/IR} logarithms. The clean separation is as follows.

\paragraph*{Separation of UV and soft IR logarithms.}
For a generic kinematic scale $Q^{2}$ (e.g. $s,t,u$ or $p^{2}$), the one–loop corrections to either the two–point function or a  scattering amplitude can be organized as
\begin{equation}
\label{eq:decomp}
\mathcal{A}(Q^{2},\Omega)
= A_{\rm UV}(\Omega)\,\ln\!\frac{Q^{2}}{\mu^{2}}
+ A_{\rm soft}(\Omega)\,\ln\!\frac{Q^{2}}{m^{2}}
+ \text{finite},
\end{equation}
where $\Omega$ collectively denotes angular variables, $\mu$ is the MS-UV regulation scale, and  $m$ is an IR regulator (here taken as a small mass). The coefficient $A_{\rm UV}$ is \emph{UV in origin}: it is gauge independent, process independent, and controls the renormalization of the charge. In contrast, $A_{\rm soft}(\Omega)$ encodes soft/collinear physics: it depends on the IR regulator (and cancels only after including processes involving emission of soft particles), can be angular dependent, and may be gauge dependent.
By construction, it must not be interpreted as a contribution to the UV (physical) running.

\paragraph*{QED vs.\ quadratic gravity.}
In QED, the vacuum polarization contains a $\ln(Q^{2}/\mu^{2})$ term whose coefficient fixes the well–known one–loop beta function; soft logarithms cancel between the diagrams. When quadratic–gravity corrections are added to the photon self–energy, we find that the combination $B_{0}(0,m^2,m^2)-B_{0}(Q^2,m^2,m^2)$
is UV finite: the $\ln\mu^2$ pieces cancel, leaving only IR–controlled logarithms such as $\ln(Q^2/m^2)$ (and constants) which can carry gauge–parameter dependence. Therefore, at one loop there is no gravitational contribution to $A_{\rm UV}$ and hence no modification of $\beta(e)$ from the quadratic–gravity sector. The example presented here emphasizes that the mere presence of $\ln(-p^2)$ does not by itself signal UV running: one must distinguish UV $\ln(Q^2/\mu^2)$ from IR $\ln(Q^2/m^2)$ pieces. In particular, dropping momentum–independent pieces such as $B_{0}(0,m^2,m^2)$ too early, can obscure cancellations that render combinations like $B_{0}(0,m^2,m^2)-B_{0}(Q^2,m^2,m^2)$ UV finite; such terms should be retained until the UV/IR separation is made explicit.

Our analysis clarifies how UV and soft logarithms should be handled when defining a ``physical'' running  in the presence of gravitational interactions. At one loop, quadratic–gravity does not modify the UV running of the electromagnetic coupling, while it can induce IR–sensitive, gauge–dependent logarithms that should not be confused with the renormalization–group evolution. A consistent treatment of Eq.~\eqref{eq:decomp} thus preserves the equivalence between the $\mu$–running and the physical running prescriptions for $\beta(e)$.

We conjecture that the apparent discrepancy in the running of gravitational couplings reported in Refs.~\cite{Buccio:2024hys,Buccio:2024omv} may originate from the premature omission of UV–divergent integrals and the consequent identification of the “physical running’’ with the logarithmic dependence of the scalar integral $B_0(p^{2},m^{2},m^{2})$, without implementing the required UV cancellations. Once the relevant UV–finite combinations are retained—e.g., $B_{0}(0,m^2,m^2)-B_{0}(p^2,m^2,m^2)$—the remaining logarithms are soft and IR–controlled. We are currently investigating this issue and will report our findings in due course. 

\acknowledgments

The work of ACL was partially supported by CNPq, Grants No.~404310/2023-0 and No.~301256/2025-0. The work of AJS has been partially supported by the CNPq project No. 309915/2021-0.

\appendix
\section*{Integrals notation}

The integrals appearing in the text are defined as follows
\begin{eqnarray}
	A_0(m^2)&=&\frac{{\mu^{2\epsilon}}}{i\pi^{2}}\int\frac{d^Dk}{(k^2-m^2)}=\frac{m^2}{\epsilon} + m^2\,\ln\left(\frac{\mu^2}{\pi m^2\,\mathrm{e}^{\gamma_E-1}}\right),\label{inta0}\\
    B_0(0,m^2,m^2)&=&\frac{{\mu^{2\epsilon}}}{i\pi^{2}}\int\frac{d^Dk}{(k^2-m^2)^2}=\frac{1}{\epsilon}+\ln\left(\frac{\mu^2}{\pi m^2\,\mathrm{e}^{\gamma_E}}\right),
    \label{intb00}\\
	B_0(p^2,m^2,m^2)&=&\frac{{\mu^{2\epsilon}}}{i\pi^{2}}\int\frac{d^Dk}{(k^2-m^2)((k-p)^2-m^2)}\nonumber\\
	&=&\frac{1}{\epsilon}+\left[ \frac{\sqrt{p^2(p^2-4m^2)}~\ln\left(\frac{\sqrt{p^2(p^2-4m^2)}+2m^2-p^2}{2m^2} \right)}{p^2}+\ln\left(\frac{\mu^2}{\pi m^2\,\mathrm{e}^{\gamma_E-2}}\right)
    \right]\nonumber\\
    &=& \frac{1}{\epsilon}+\ln\left(-\frac{\mu^2}{\pi \, \mathrm{e}^{\gamma_E-2}\, p^2} \right) +\mathcal{O}(m^2/p^2),\label{intb0}
\end{eqnarray}
\noindent where in the last step of Eq.\eqref{intb0} we have expanded it for $p^2\gg m^2$, $D=4-2\epsilon$ and $\gamma_E$ is the Euler-Mascheroni constant.

In particular, in dimensional regularization (Eqs. \eqref{inta0} and \eqref{intb00}) one may express the zero–momentum two–point function in terms of the tadpole as
\begin{equation}
m^2 B_{0}(0,m^2,m^2)
=A_{0}(m^2)- m^2.
\end{equation}

To illustrate that UV–finite loop functions can nevertheless carry momentum logarithms dominated by IR physics, consider the scalar integral
\begin{eqnarray}
C_{0}\!\left(0,p^{2},p^{2};m^{2},m^{2},m^{2}\right)
&=&\frac{\mu^{2\epsilon}}{i\pi^{2}}\int\!\frac{d^{D}k}{\big(k^{2}-m^{2}\big)\big[(k-p)^{2}-m^{2}\big]^{2}}\nonumber\\
&=&\frac{\ln\!\left(\dfrac{\sqrt{p^{2}\!\left(p^{2}-4m^{2}\right)}+2m^{2}-p^{2}}{2m^{2}}\right)}
{\sqrt{p^{2}\!\left(p^{2}-4m^{2}\right)}}+\mathcal{O}(\epsilon)
\nonumber\\
&=& \frac{\ln\!\left(-\dfrac{p^2}{m^2}\right)}{p^{2}}
+\mathcal{O}\!\left(\frac{m^{2}}{p^{2}}\right),
\label{C0soft}
\end{eqnarray}
which is UV finite and yet exhibits a $\ln(-m^2/p^2)$ dependence at large momentum, entirely controlled by the IR regulator $m$. Similarly, the UV–finite combination of two–point functions
\begin{equation}
B_{0}\!\left(p^{2},m^{2},m^{2}\right)-B_{0}\!\left(0,m^{2},m^{2}\right)
 \propto \ln\frac{-m^2}{p^2} + \text{constants},
\label{B0diffsoft}
\end{equation}
contains only soft (nonlocal) logarithms and does not contribute to the UV coefficient that determines the physical running of the coupling. These examples emphasize that the mere presence of $\ln(-p^{2})$ does not by itself signal UV running: one must distinguish UV $\ln(Q^{2}/\mu^{2})$ from IR $\ln(Q^{2}/m^{2})$ pieces. In particular, dropping momentum–independent pieces such as $B_{0}(0,m^{2},m^{2})$ too early can obscure cancellations that render combinations like \eqref{B0diffsoft} UV finite; such terms should be retained until the UV/IR separation is made explicit.

\vspace{1cm}

\begin{thebibliography}{99}

\bibitem{'tHooft:1974bx} 
G.~'t Hooft and M.~J.~G.~Veltman,
Ann. Inst. Henri Poincare A {\bf 20}, 69 (1974).

\bibitem{PhysRevLett.32.245} 
S.~Deser and P.~van Nieuwenhuizen,
Phys. Rev. Lett. {\bf 32}, 245 (1974).

\bibitem{Deser:1974cy} 
S.~Deser and P.~van Nieuwenhuizen,
Phys. Rev. D {\bf 10}, 411 (1974).

\bibitem{Donoghue:1994dn} 
J.~F.~Donoghue,
Phys. Rev. D {\bf 50}, 3874 (1994).

\bibitem{Burgess:2003} 
C.~P.~Burgess,
Living. Rev. Relativity {\bf 7}, 5 (2004).

\bibitem{Donoghue:2012zc} 
J.~F.~Donoghue,
AIP Conf. Proc.  {\bf 1483}, 73 (2012).

\bibitem{Donoghue:2017pgk}
J.~F.~Donoghue, M.~M.~Ivanov, and A.~Shkerin,
arXiv:1702.00319.

\bibitem{Faller:2007sy}
S.~Faller,
Phys. Rev. D \textbf{77}, 124039 (2008).

\bibitem{Bjerrum-Bohr:2014zsa}
N.~E.~J.~Bjerrum-Bohr, J.~F.~Donoghue, B.~R.~Holstein, L.~Plant\'e and P.~Vanhove,
Phys. Rev. Lett. \textbf{114}, no.6, 061301 (2015).

\bibitem{Donoghue:2014yha}
J.~F.~Donoghue and B.~K.~El-Menoufi,
Phys. Rev. D \textbf{89}, no.10, 104062 (2014).

\bibitem{Rovelli:1997yv}
C.~Rovelli,
Living Rev. Relativity \textbf{1}, 1 (1998).

\bibitem{Percacci:2007sz}
R.~Percacci,
arXiv:0709.3851.

\bibitem{Hamber:2009mt}
H.~W.~Hamber,
Gen. Rel. Grav. \textbf{41}, 817 (2009).

\bibitem{polchinski}
J. Polchinski, \textit{String Theory}, Vols. 1 and 2
(Cambridge University Press, Cambridge, England, 1998).

\bibitem{Surya:2019ndm}
S.~Surya,
Living Rev. Rel. \textbf{22}, no.1, 5 (2019)

\bibitem{Stelle:1976gc}
K.~S.~Stelle,
Phys. Rev. D \textbf{16}, 953-969 (1977).

\bibitem{Avramidi:1985ki}
I.~G.~Avramidi and A.~O.~Barvinsky,
Phys. Lett. B \textbf{159}, 269-274 (1985).

\bibitem{deBerredo-Peixoto:2003jda}
G.~de Berredo-Peixoto and I.~L.~Shapiro,
Phys. Rev. D \textbf{70}, 044024 (2004).

\bibitem{Narain:2013eea}
G.~Narain and R.~Anishetty,
JHEP \textbf{10}, 203 (2013).

\bibitem{Salvio:2014soa}
A.~Salvio and A.~Strumia,
JHEP \textbf{06}, 080 (2014).

\bibitem{Alvarez-Gaume:2015rwa}
L.~Alvarez-Gaume, A.~Kehagias, C.~Kounnas, D.~L\"ust and A.~Riotto,
Fortsch. Phys. \textbf{64}, no.2-3, 176-189 (2016).

\bibitem{Salvio:2017qkx}
A.~Salvio and A.~Strumia,
Eur. Phys. J. C \textbf{78}, no.2, 124 (2018).

\bibitem{Holdom:2015kbf}
B.~Holdom and J.~Ren,
Phys. Rev. D \textbf{93}, no.12, 124030 (2016).

\bibitem{Anselmi:2018ibi}
D.~Anselmi and M.~Piva,
JHEP \textbf{05}, 027 (2018).

\bibitem{Salvio:2018crh}
A.~Salvio,
Front. in Phys. \textbf{6}, 77 (2018).

\bibitem{Salvio:2018kwh}
A.~Salvio, A.~Strumia and H.~Veerm{\"a}e,
Eur. Phys. J. C \textbf{78}, no.10, 842 (2018).

\bibitem{Donoghue:2021cza}
J.~F.~Donoghue and G.~Menezes,
Nuovo Cim. C \textbf{45}, no.2, 26 (2022).

\bibitem{Buccio:2024hys}
D.~Buccio, J.~F.~Donoghue, G.~Menezes and R.~Percacci,
Phys. Rev. Lett. \textbf{133}, no.2, 021604 (2024). 

\bibitem{Holdom:2021oii}
B.~Holdom,
JHEP \textbf{04}, 133 (2022).

\bibitem{Holdom:2021hlo}
B.~Holdom,
Phys. Rev. D \textbf{105}, no.4, 046008 (2022).

\bibitem{Cunha:2025djg}
I.~F.~Cunha and A.~C.~Lehum,
Phys. Rev. D \textbf{112}, no.2, 025016 (2025).

\bibitem{Buccio:2024omv}
D.~Buccio, L.~Parente and O.~Zanusso,
Phys. Rev. D \textbf{111}, no.6, 065022 (2025).

\bibitem{Buccio:2025tci}
D.~Buccio, G.~P.~De Brito and L.~Parente,
[arXiv:2507.15835 [hep-th]].

\bibitem{Salvio:2025cmi}
A.~Salvio, A.~Strumia and M.~Vitti,
[arXiv:2507.08803 [hep-th]].

\bibitem{Choi:1994ax} 
  S.~Y.~Choi, J.~S.~Shim and H.~S.~Song,
  Phys.\ Rev.\ D {\bf 51}, 2751 (1995). 

\bibitem{feyncalc} 
  R. Mertig and M. Bahm and A. Denne,
  Comp. Phys. Comm. {\bf 64}, 345 (1991);
  V.~Shtabovenko, R.~Mertig and F.~Orellana,
Comput. Phys. Commun. \textbf{207}, 432-444 (2016);
V.~Shtabovenko, R.~Mertig and F.~Orellana,
Comput. Phys. Commun. \textbf{256}, 107478 (2020).

\bibitem{feynarts} 
T.~Hahn,
Comput. Phys. Commun. \textbf{140}, 418-431 (2001).

\bibitem{feynrules}
N.~D.~Christensen and C.~Duhr,
Comput. Phys. Commun. \textbf{180}, 1614-1641 (2009);
A.~Alloul, N.~D.~Christensen, C.~Degrande, C.~Duhr and B.~Fuks,
Comput. Phys. Commun. \textbf{185}, 2250-2300 (2014).

\bibitem{feynhelpers}
V.~Shtabovenko,
Comput. Phys. Commun. \textbf{218}, 48-65 (2017).

\bibitem{tHooft:1978jhc}
G.~'t Hooft and M.~J.~G.~Veltman,
Nucl. Phys. B \textbf{153}, 365-401 (1979).

\bibitem{Peskin:1995ev}
M.~E.~Peskin and D.~V.~Schroeder,
``An Introduction to quantum field theory,''
Addison-Wesley, 1995,
ISBN 978-0-201-50397-5, 978-0-429-50355-9, 978-0-429-49417-8
doi:10.1201/9780429503559

\bibitem{Donoghue:2024uay}
J.~F.~Donoghue,
[arXiv:2412.08773 [hep-th]].

\bibitem{Zwanziger:1974jz}
D.~Zwanziger,
Phys. Rev. D \textbf{11}, 3481 (1975).

\bibitem{Schwartz:2014sze}
M.~D.~Schwartz,
\textit{Quantum Field Theory and the Standard Model}
(Cambridge University Press, Cambridge, England, 2014). 

\end{thebibliography}
\end{document}